\documentclass[11pt]{article}
\usepackage{amsmath,amsfonts,graphics,epsfig}
\usepackage{latexsym}
\usepackage{graphicx}
\usepackage{amssymb}
\usepackage{amsmath}
\usepackage{amsfonts}
\usepackage{color}

\newcommand{\be}{\begin{equation}}
\newcommand{\ee}{\end{equation}}

\begin{document}
\newtheorem{thm}{Theorem}[section]
\newtheorem{prop}{Proposition}[section]
\newtheorem{lem}{Lemma}[section]
\newtheorem{cor}{Corollary}[section]
\newtheorem{tdef}{Definition}[section]
\newtheorem{rem}{Remark}[section]
\renewcommand{\theequation}{\thesection.\arabic{equation}}
\newcommand{\tsquare}{\hfill \rule{3mm}{3mm}}

\def\rit{\hbox{\rm I\hskip -2pt R}}
\def\nit{\hbox{\rm I\hskip -2pt N}}
\def\cit{\hbox{\rm l\hskip -5.5pt C\/}}
\def\zit{\hbox{\rm Z\hskip -4pt Z}}
\def\qit{\hbox{\rm l\hskip -5.5pt Q}}
\def\notin{\hbox{$\in \hskip -10pt / \; $}}
\def\ind{\hbox{${\rm 1\hskip -3pt I}$}}
\def\mean{\hbox{${\rm I \hskip -2pt E}$}}
\def\Pr{\hbox{${\rm I \hskip -2pt P}$}}
\def\cond{\hbox{$ \stackrel{\cal D}{\rightarrow}$}}
\def\condas{\hbox{$\stackrel{\pr-a.s.}{\rightarrow}$}}
\def\dint{\hbox{$\int_{-\infty}^\infty$}}
\def\endpf{\hbox{${\tsquare}$}}

\begin{center}
{\Large {\bf On Powers of Gaussian White Noise}}
\end{center}
\vspace{1.5cm}

\begin{center}
{ \bf A. V. BALAKRISHNAN \footnotemark[1] and Ravi R. MAZUMDAR\footnotemark[2]
}

\end{center}
\vspace{1cm}
\noindent \footnotemark[1] Departments of Electrical Engineering and Mathematics, University of California, Los Angeles, Ca. 90024, USA

\noindent \footnotemark[2] Department of Electrical and Computer Engineering, University of Waterloo, Waterloo, ON N2L 3G1, Canada
\vspace{1cm}

\begin{center}
\today
\end{center}
\vspace{0.5cm}

\begin{abstract}
Classical Gaussian white noise in communications and signal processing is viewed as the limit of  zero mean second order Gaussian processes with a compactly supported flat spectral density as the support goes to infinity. The difficulty of developing a theory to deal with nonlinear transformations of white noise has been to interpret the corresponding limits. In this paper we show that a renormalization and centering of powers of band-limited Gaussian processes is Gaussian white noise and as a consequence, homogeneous polynomials under suitable renormalization remain white noises. 
\end{abstract}
\vspace{0.5cm}

Asymptotics
\vspace{0.3cm}

%\noindent{\bf AMS Classification: 28C20, 60G30} 
\vspace{0.5cm}

\noindent{\bf Short-title:} Powers of white noise
\vspace{0.5cm}

\newpage
\section{Introduction}

White noise is  critical in the development of statistical signal processing and in models for communication channels. The centrality of the process is that any second order covariance function can be realized as the result of white noise through a linear filter. In the stationary case white noise is the basic building block of of constructing optimal filters. In the classical context, white noise is viewed as a limit of a second order process that has a flat spectral density of compact or finite support (referred to as the bandwidth of the process) as the support becomes infinite \cite{WH1983,Hida1967,AVB1976a}. Such processes cannot be physically realizable because they would have infinite energy, and yet white noise plays a crucial role in developing practical filters.

The difficulty of defining white noise is not just because of the infinite energy. Indeed defining a continuous-time white noise process presents difficulty even in the probabilistic context. This is because the induced probability measure cannot be countably additive on the space $L_2[0,T]$  i.e. constructing a Gaussian process whose covariance function is a Dirac delta function results in the underlying probability measure being only finitely additive,  and thus the white noise map $n_t (\omega)$ is not a bona fide random variable \cite{AVB1980,Saz,Gel}. 
Indeed because of the difficulty of mathematically dealing with white noise, Balakrishnan in a series of papers \cite{AVB1974,AVB1976a,AVB1981} developed a finitely additive framework for analyzing white noise processes and the associated calculus. This was further developed through the idea of liftings in the work of Kallianpur and Karandikar \cite{KK1988}. An alternative approach exploiting the structure of abstract Wiener spaces can be found in the work of Kuo \cite{Kuo1975}, and Gross \cite{Gross1962,Gross1963} for example, where the idea is to work through on the lifted space through the lifting map that results working on the abstract Wiener space.

 Classical white noise is only  defined as a generalized process, i.e., it does not induce a countably additive measure on the Hilbert space. However certain transformations of white noise do induce countable additive measures.  This is closely related to the result of Sazonov \cite{Saz} on the existence of countable additive measures on Hilbert spaces. In the linear context, any continuous linear operator such as a kernal operator acting on white noise results in the resulting map defining a bona fide stochastic process.  The integral operator $Ln(t) = \int_0^t n_sds$ as a mapping from $L_2[0,T] \rightarrow L_2[0,T]$ induces a countably additive measure whose extension to $C(0,T)$ \cite{ AVB1981, Germani1984} is the Wiener measure \cite{Kuo1975,Smol1973}. Viewed this way, Gaussian white noise is the derivative of the Wiener process even though the Wiener process is not differentiable almost surely for any t. However when restricting ourselves to linear operators we can make sense because in most applications (filtering and communications) such operators are Hilbert-Schmidt \cite{AVB1981,Kuo1975} and we can interpret the results both probabilistically as well as algorithmically.  For nonlinear transformations the situation is more complicated. Indeed one class of nonlinear transformations that induce countably additive measures is that associated with S-continuous mappings where the S-topology is that which is associated with semi-norms associated with nuclear operators \cite{AVB1981, BM1993, BM1994, GG83}.

There still remains the question whether nonlinear transformations, the powers for example, of white noise can be suitably interpreted, even as generalized processes \cite{Gikh,Gel}. The need for such interpretation can be found in many applications in mathematical and quantum physics \cite{accardi2002, San1981} for example. Indeed, in \cite{San1981} the authors provide a heuristic justification for treating a renormalization of a squared white noise term as white noise.

In this paper we show that under a suitable renormalization, integral powers of Gaussian white noise viewed as the limit of a band-limited Gaussian process with flat spectral density is indeed Gaussian white noise, a  non-trivial fact given that non-linear transformations of Gaussian random variables are not Gaussian.

\section{Some preliminaries}

Let us first see some of the classical results related to white noise. For simplicity we restrict ourselves to real-valued processes. The extension to $\Re^d$ valued processes is direct. We denote the inner product in $L_2[0,T]$ by $[f,g]_T= \int_0^T f(s)g(s)ds, \ f,g \in L_2[0,T]$ and the norm is denoted by $[f,f]_T = ||f||_T^2$. The space $L_2(-\infty,\infty)$ is simply denoted as $L_2$

Let $\{X_W(t),\ -\infty < t < \infty\}$ denote a  zero mean stationary Gaussian process. Let $R(t) = \mean[X(t+s)X(s)$ denote the covariance and it is assumed that $\int_{-\infty}^{\infty} |R(t)|dt < \infty$.
By Bochner's theorem there exists a spectral density $S(\lambda), -\infty < \lambda , \infty$ and $S(\lambda) = \int_{-\infty}^{\infty} R(t) e^{-i2\pi \lambda t}dt$.

Now suppose $S(\lambda)$ is band-limited and flat as follows:
\begin{eqnarray*}
S_W(\lambda) & = & 1 \ \ , -W\leq \lambda < W \\
& = & 0\ \ {\rm otherwise}
\end{eqnarray*}

Now from the fact that $S(\lambda)$ has support in $[-W,W]$ it follows that there exists a spectral process that is Gaussian and independent on non-overlapping intervals denoted by  $\hat{X}_W(\lambda)$ such that for every $(a,b), \ \ \int_a^bS_W(\lambda)d\lambda = \mean[\int_a^b \int_a^b \hat{X}(d\lambda)\hat{X}(d\lambda')]$ and 
\be
X(t) = \int_{-\infty}^{\infty} e^{i2\pi\lambda t} \hat{X}_W(\lambda) d\lambda,\ \ -\infty < t  <\infty
\ee

Moreover the limit in mean square (denoted by q.m) 
\begin{equation}
\lim_{W\to\infty} X_W(t)\qquad{ {{q.m}\atop=} } \qquad X(t) 
\end{equation}
exists and is called Gaussian white noise.

In particular we see that such a limiting process process will have the following properties:

\begin{description}
\item{i)} The random variables $y= \int_0^T X(s)f(s)ds$ will be distributed as $N(0, ||f||_T^2)$ where $f\in L_2[0,T]$.
\item{ii)} $\mean\left([f,X]_T[g,X]_T\right)= [f,g]_T$
\item{iii)} Let $\phi_i(t)$ be an orthonormal functions in $L_2[0,T]$ and consider the collection of random variables $Y= col(y_1,y_2,\ldots,y_N)$ where $y_i= \int_0^T\phi_i(t)X(t)dt$. Then 
$Y \sim N ({\bf 0}, I_N)$ where ${\bf 0}$ is the N-dimensional vector of all 0's and $I_N$ is the $N\times N$ identity matrix.
\item {iv)} $\mean[\int_0^T f(s) X(s)ds X(t)]= f(t), 0 \leq t \leq T$ and 0 otherwise.
\end{description}

This is equivalent to saying that $\{X(t),\ -\infty < t < \infty\}$ is a zero mean stationary Gaussian process with covariance $R(t,s) = R(t-s) = \delta(t-s)$ where $\delta(.)$ denotes the Dirac delta function.

Indeed let  \be
R_W(t)= \int_{-W}^W S_W(\lambda)e^{i2\pi\lambda}d\lambda= \frac{\sin 2\pi Wt}{\pi t},\ -\infty < t < \infty
\ee
and hence for any $f\in L_2$ we have:
\be
\lim_{W\to\infty} \int_{-\infty}^{\infty} R_W(t-s)f(s)ds = f(t)
\ee
where the limit is in $L_2$.

Clearly such a process is not physically realizable since by Bochner's theorem $R(0) = \lim_{W\to\infty} \int_{-W}^W 1. d\lambda = \infty$. in other words its sample paths cannot be in $L_2$. It is worth remarking that from above: the process $Y(t)= \lim_{W\to\infty}\int_0^t X_W(s)ds$ is a zero mean Gaussian process with variance $t$ or is Brownian motion. The point is that the process $X(t)$ is not well defined and thus $X(t)$ only formally the derivative of $Y(t)$.

Herein lies the problem. Clearly we can make sense of operations when $L_2$  functions act on $X$ and in hence problems where white noise is the input to a linear time-invariant system we can give mathematical meaning by the limiting arguments (in q.m). However even simple nonlinear operations such as squaring, i.e., $Y(t)= X^2(t)$ do not make sense because such a process would have infinite mean and its covariance would be the product of delta functions that is not defined in any meaningful way.

In the following section we show that the squaring operation does make sense if we perform a suitable renormalization of the process $X_W(.)$ and then the limiting process itself is Gaussian white noise. This is indeed an unexpected result because squaring a simple Gaussian random variable results in a chi-squared random variable.

\section{Renormalized powers of white noise}

First note that from the definition of $\{X_W(t)\}, -\infty < t < \infty$ is a stationary Gaussian process, we can represent $X_W(t)$ as:
\be
\label{proj}
X_W(t) = \frac{R_W(t)}{R_W(0)}X(0) + \nu_W(t)
\ee
where $\nu_W(t)$ is a zero mean Gaussian r.v. independent of $X(0)$ and variance given by:
\be
\mean[\nu^2_W(t)]= R_{\nu,W}(t) = \frac{R^2_W(0)- R^2_W(t)}{R_W(0)}
\ee

Let us now define the following process:
\be
\label{squareproc}
Y_W(t) = \frac{1}{2\sqrt{W}}(X^2_W(t) - \mean[X^2_W(t)]) = \frac{X^2_W(t) - 2W}{2\sqrt{W}},\ \ -\infty < t < \infty
\ee

Then $Y_W(t)$ is a centered (mean 0) , renormalized process that denotes the nonlinear transformation (squaring) of the pre-white noise process. We now state and prove the main result.

\begin{thm}
\label{squarethm}
Consider the renormalized and centered process $Y_W(t)$ defined in \ref{squareproc} above. Then:
\be
\lim_{W\to\infty} Y_W(t) \ \  = \ \ Y(t)\ in\ L_2(\Pr)\times L_2
\ee
Moreover $Y(t)$ is Gaussian white noise.
\end{thm}

We prove the result through the following two results.

\begin{prop}
\label{deltaprop}
Let $R_W^Y(t)$ denote the covariance of $Y_W(t)$. Then for every $f(.) \in L_2$
\be
\lim_{W\to \infty} \int_{-\infty}^\infty  R_W^Y(t-s) f(s)ds = f(t),\ \ -\infty < t < \infty
\end{equation}
or formally:
$$\lim_{W\to\infty} R^Y_W(t) = \delta(t)\qquad{ -\infty < t < \infty}$$
where $\delta(.)$ is the Dirac delta function.
\end{prop}
\vspace{0.5cm}

\noindent{\bf Proof:} 

First note that  $Y_W(t)$ is a w.s.s. process whose covariance denoted by 
$$R_W^Y(t)= \frac{1}{2W}R^2_W (t)= \frac{1}{2W}\left(\frac{\sin 2\pi Wt}{\pi t}\right)^2$$

Next note that  by Parseval's theorem:
$$\frac{1}{2W}\int_{-\infty}^{\infty} \left(\frac{\sin 2\pi Wt}{\pi t}\right)^2dt = \frac{1}{2W} \int_{-\infty}^{\infty} \ind^2_{[-W,W]}(\lambda)d\lambda= 1$$
Define the measure $d\phi_W(t) = \frac{1}{2W} \left(\frac{\sin 2\pi Wt}{\pi t}\right)^2dt $ so that $\phi_W(-\infty,\infty)=1= \int_{-\infty}^{\infty}\phi_W(t)dt$.
Hence for any continuous $f(.) \in L_2$ we have:
\begin{eqnarray*}
\dint |\dint f(t+s)d\phi_W(s)- f(t)|^2 dt & = & \dint |\dint (f(t+s)-f(t))d\phi_W(s) |^2 dt\\
& \leq & \dint \dint | f(t+s)-f(t)|^2 d\phi_W(s) dt 
\end{eqnarray*}
where we have used Minkowski's \cite{Hardy1934} inequality for integrals in the second step noting that $d\phi_W(.)$ defines a measure that puts mass 1 on $(-\infty,\infty)$.
Now performing a change of variables by substituting $2\pi W s= \tau$ we can re-write:
\begin{eqnarray*}
 \dint \dint |f(t+s)-f(t)|^2 d\phi_W(s) dt & = &\dint \dint |f(t + \frac{\tau}{2\pi W})- f(t)|^2 \frac{1}{\pi}\left(\frac{ \sin \tau}{\tau}\right)^2d\tau\\
 & \leq & \frac{1}{\pi} ||g_W(t,\tau)(\frac{\sin \tau}{\tau})||^2_{L_2\times L_2}
 \end{eqnarray*}
where $g_W(t,\tau)= (f(t+\frac{\tau}{2\pi W})-f(t))$ and $L_2\times L_2 = L_2(-\infty,\infty)\times L_2(-\infty,\infty)$ with the product measure defined thereon.

Now , for every fixed $\tau$
$$\dint |g_W(t,\tau)|^2 dt \leq 2 ||f||^2$$
and hence using Fubini's theorem and the fact that $\dint (\frac{\sin \tau}{\tau})^2 d\tau = \pi $
$$\dint \dint |g_W(t,\tau)(\frac{\sin \tau}{\tau})|^2 dt d\tau \leq 2\pi ||f||^2$$
Moreover $|g_W(t,\tau)| \to 0\ as\ W\to \infty$ for every $\tau$ fixed we have that 
$$\dint|\dint (f(t+s)\phi_W(s)ds -f(t)|^2 dt \rightarrow 0\ as \ W\rightarrow \infty$$
by dominated convergence.

$$\eqno{\tsquare}$$
\vspace{0.5cm}

The second result we prove is the convergence to a Gaussian process. For this we need the following result.

\begin{lem}
Let $\{X_W(t),\ -\infty < t < \infty\}$ be a zero mean stationary Gaussian process with bandlimited spectral density. Let $(a,b)$ and $(c,d)$ be any non-overlapping intervals in $\Re$. Let $h(.) \in L_2$. Define the random variables $X^W_{a,b}$ (resp. $X^W_{cd})$ as $X^W_{a,b} = \int_a^b X^W(s)h(s)ds$ (similarly for $X^W_{c,d}$).

Then $(X^W_{ab},X^W_{c,d})$ are asymptotically independent as $W\rightarrow \infty$
\end{lem}

\noindent{\bf Proof:} To show the result it suffices to show that the random variables are asymptotically uncorrelated since they are jointly Gaussian by construction.

$$\mean[X^W_{ab}X^W_{cd}]  =  \int_{-\infty}^\infty \int _{-\infty}^\infty \ind_{(a,b)}(u)\ind_{(c,d)}(v) h(u)h(v)R^W(u-v)dudv$$

Hence from applying the result of Proposition \ref{deltaprop} we have:
$$\lim_{W\to\infty} \mean[X^W_{a,b}X^W_{cd}] = \int_{-\infty}^\infty \ind_{(a,b)}(u)\ind_{(c,d)}(u)h(u)h(v)du = 0$$
since $(a,b)$ and $(c,d)$ are non-overlapping.

\vspace{0.5cm}
\begin{rem}
From above it readily follows that the random variables $Y^W_{ab}$ and $Y^W_{cd}$ defined in an analogous way are also asymptotically independent  as $W\to \infty$ since they are functionals of the underlying process $X^W_.$ on non-overlapping intervals.
\end{rem}
\vspace{0.3cm}

Let $C_W(h)$ denote the characteristic functional of $\{Y^W_t\}, -\infty < t < \infty$ defined as:
\be
C_W(h) = \mean[ e^{i[Y^W,h]_T}],\ \ h\in L_2(0,T)
\ee
and $[x,y]_t= \int_0^t x(s)y(s)ds$ for $x, y \in L_2[0,T],\qquad t\leq T$ .

\begin{prop}
\label{gaussian}
\be
\lim_{W\to\infty} C_W(h) = e^{-\frac{1}{2}||h||^2_T}
\ee
or $\{Y^W_t\}$ converges to Gaussian white noise in $L_2(\Pr)\times L_2$

\end{prop}
\vspace{0.3cm}

\noindent{\bf Proof}

First note that from the asymptotic independence we have:
$$\lim_{W\to\infty} \mean[e^{i[Y^W,h]_{t+s}}] = \lim_{W\to\infty} \mean[e^{i[Y^W,h]_t}] \mean[e^{i[Y^W,h]^{t+s}_t}]$$
where $[x,y]_a^b= \int_a^b x(s)y(s)ds$.

Now to show that $\{Y^W_t\}$ converges to Gaussian white noise it is sufficient to show that:
\be
\lim_{W\to\infty} \frac{\frac{d}{dt}\mean[ e^{i[Y^W,h]_t}]}{\mean[e^{i[Y^W,h]_t}]} = -\frac{1}{2} h^2(t)\qquad a.e.\ in\ t
\ee

Note that from the boundedness of $e^{i[Y^W,h]_t}$ and the convergence in q.m. of $Y^W_.$ it is easy to show that:
$$\lim_{W\to\infty} \frac{d}{dt} \mean[e^{i[Y^W,h]_t}]= \frac{d}{dt} \lim_{W\to\infty} \mean[e^{i[Y^W,h]_t}]$$

Now:
\be
\label{deriv}
\frac{d}{dt} \mean[e^{i[Y^W,h]_t}] = \lim_{\Delta \to 0} \frac{1}{\Delta} \mean\left[ e^{i[Y^W,h]_{t+\Delta}}] -e^{i[Y^W,h]_t}]\right]
\ee
where the limit on the r.h.s. of  (\ref{deriv}) is to be interpreted in $L_2(P)$.

Hence:
$$
\frac{1}{\Delta} \mean\left[ e^{i[Y^W,h]_{t+\Delta}}] -e^{i[Y^W,h]_t}]\right]  =  \frac{1}{\Delta} \mean[e^{i[Y^W,h]_t} \left( e^{i\xi^W (\Delta)} - 1\right)]\\
$$

where $\xi^W(\Delta) = [Y^W,h]^{t+\Delta}_t$.

Now we make use of the following identity:
$$e^{i x} = 1 + ix -\frac{1}{2}x^2 -\frac{1}{2}\int_0^x s^2e^{i(x-s)}ds$$
to obtain for each $\Delta >0$ by asymptotic independence:
$$\lim_{W\to \infty} \mean[e^{i[Y^W,h]_t}\xi^W(\Delta)] = \lim_{W\to \infty} \mean[e^{i[Y^W,h]_t}] \mean[\xi^W(\Delta)] = 0$$

For the second term we have:
$$\lim_{W\to\infty} \mean[e^{i[Y^W,h]_t}\left(\xi^W(\Delta)\right)^2] = \lim_{W\to\infty} \mean[e^{i[Y^W,h]_t}]\mean\left(\xi^W(\Delta)\right)^2$$
Now from Proposition \ref{deltaprop} we have:
$$\lim_{W\to \infty} \mean \left(\xi^W(\Delta)\right)^2 = \int_t^{t+\Delta} h^2(u)du$$

For the third term we obtain:
\begin{eqnarray*}
|\mean[e^{i[Y^W,h]_t} \int_0^{\xi^W(\delta)} s^2 e^{i(\xi^W(\Delta)-s)}ds]| & \leq & \mean |\xi^W(\Delta)|^3 \\
& \leq  & \left( \mean[\xi^W(\Delta)|^2\right)^{\frac{3}{2}}
\end{eqnarray*}
Therefore, once again using the result of Proposition \ref{deltaprop} we obtain:
$$\lim_{W\to\infty} \left( \mean[\xi^W(\Delta)|^2\right)^{\frac{3}{2}} = \left(\int_t^{t+\Delta} h^2(u)du\right)^{\frac{2}{2}}  = O(\Delta^{\frac{3}{2}})$$

Therefore combing all the three estimates above we obtain 
\begin{eqnarray*}
\lim_{\Delta\to 0}\lim_{W\to\infty} \frac{ \frac{1}{\Delta} \mean\left[ e^{i[Y^W,h]_{t+\Delta}}] -e^{i[Y^W,h]_t}]\right]
}{\mean[e^{i[Y^W,h]_t}]} & = & \lim_{\Delta \to 0} \frac{1}{\Delta}\left (\int_t^{t+\Delta} h(u)^2du + O(\Delta^{\frac{3}{2}})\right)\\
& = & h^2(t)\ \ a.e. t
\end{eqnarray*}

Therefore we obtain:
$$\lim_{W\to\infty} \mean[e^{i[Y^W,h]_T}] = e^{-\frac{1}{2} ||h||^2_T}$$
or the limiting process has a characteristic functional that corresponds to standard Gauss measure on $L_2[0,T]$ for every $T$ and hence is Gaussian white noise.

This completes the proof.

$$\eqno{\tsquare}$$

A consequence of the above result is that any integral power of white noise should remain white noise under proper re-normalization. Indeed it is the case and we show this below.

Now let $a_W(t)= \frac{R_W(t)}{R_W(0)}$, and so for any integer $n$, using (\ref{proj}) we obtain:
\be
\label{binom}
X^n_W(t) = (a(t)X_W(0) + \nu_W(t))^n = \sum_{p=0}^n {n\choose p} a_W^p(t)X_w^p(0) \nu^{n-p}_W(t)
\ee
From the independence of $X(0)$ and $\nu_W(t)$ we have:
\begin{eqnarray}
R_W^n(t ) & = & cov(X^n_W(t) X^n_W(0)] = \sum_{p=0}^n a^p_W(t) \mean[X_W^{n+p}(0)] \mean[\nu_W^{n-p}(t)]- \left(\mean[X_W^n(0)]\right)^2 \nonumber\\
& = & a^n(t)\mean[X^{2n}_W(0)] + \sum_{p=0}^{n-1} a^p_W(t)\mean[X_W^{n+p}(0)] \mean[\nu_W^{n-p}(t)]- \left(\mean[X_W^n(0)]\right)^2
\end{eqnarray}

Define:
\be
\label{higherpower}
Y^n_W(t) = \frac{X^n_W(t)- \mean[X^n_W(0)]}{\sqrt{(n-1)!!}(2W)^{\frac{n}{2}}},\ \ n\geq 2
\ee
where $(n-1)!!= (n-1)(n-3)(n-5)\ldots 1$.
\vspace{0.3cm}

Then we can state the following theorem about higher order powers of white noise.

\begin{thm}
Let $\{X_W(t)\}, -\infty < t < \infty$ be a Gaussian process whose spectral density is flat, of unit power and with support in $[-W,W]$.

Then the process $\{Y_W^n(t)\}, -\infty < t < \infty$ converges to Gaussian white noise in $L_2(\Pr)\times L_2$ as $W\rightarrow \infty$.
\end{thm}
\vspace{0.3cm}

\noindent{\bf Proof:} The proof essentially follows from the arguments in the proof of Theorem \ref{squarethm}. Indeed the result follows from the fact that for any Gaussian $N(m,\sigma^2)$ r.v. 
the moment of order $n$ is given by:
\begin{eqnarray*}
\mean[(X-m)^n] & = & (n-1)!!\sigma^n,\ \ \ \ (n-1)!!= (n-1)(n-3)(n-5)\ldots 1,\ \ n\ even\\
& = & 0,\ \ n\ odd
\end{eqnarray*}

Note covariance of $Y^n_W$ is just given by $\frac{R_W^n(t)}{(n-1)!!(2W)^{n}}$. Now from 3.15 it can be seen that $R^n_W(t)= C. R_W^n(t)$ where $C$ is a constant.

Let $f(.)\in L_2$ and consider:
$$\int_{-\infty}^{\infty} \frac{R_W^n(t-s)}{(n-1)!!(2W)^{n}}f(s)ds = \frac{1}{(n-1)!!(2W)^{n}}\int_{-\infty}^{\infty} \left(\frac{\sin 2\pi W(t-s)}{\pi (t-s)}\right)^{n}f(s)ds$$
Noting that $|R_W(t)| \leq R_W(0)$ we have : 
$$\frac{1}{(n-1)!!(2W)^{n}}|\int_{-\infty}^{\infty} \left(\frac{\sin 2\pi W(t-s)}{\pi (t-s)}\right)^{n}f(s)ds| \leq |R_W(0)|^n ||f|| \leq C_1 ||f|| < \infty$$
Furthermore, see \cite{grad} :
$$\frac{1}{(n-1)!!(2W)^{n}}\int_{-\infty}^{\infty} \left(\frac{\sin 2\pi W(t-s)}{\pi (t-s)}\right)^{n}dt = C_3(n) < \infty$$

Then, we can repeat the arguments as in Propositions \ref{deltaprop} and \ref{gaussian} {\em mutatis mutandis} to show that the normalized process is Gaussian white noise whose variance depends on the constants.
$$\eqno{\tsquare}$$

\begin{rem}
This result can be directly extended to homogeneous polynomials $P(X_W(t))$ in the obvious way by defining the renormalization factor to normalize the highest power of the polynomial. It can be shown the lower order powers do a play a role in the asymptotic limit.
\end{rem}
\vspace{0.5cm}

 \section*{Acknowledgement}

The research of RM has been supported in part by a grant from the Natural Sciences and Engineering Research Council of Canada (NSERC) through the Discovery Grant program. He would also like to thank Patrick Mitran for useful discussions.

\bibliographystyle{plain}

\bibliography{whitenoise}

\end{document}